\documentclass{svproc}

\usepackage{mathptmx}       % selects Times Roman as basic font
\usepackage{helvet}         % selects Helvetica as sans-serif font
\usepackage{courier}        % selects Courier as typewriter font
\usepackage{type1cm}        % activate if the above 3 fonts are
\usepackage[ruled,vlined,linesnumbered]{algorithm2e}

\usepackage{makeidx}         % allows index generation
\usepackage{graphicx}        % standard LaTeX graphics tool
\usepackage{multicol}        % used for the two-column index
\usepackage[bottom]{footmisc}% places footnotes at page bottom

\usepackage{array,multirow,makecell}
\setcellgapes{1pt}

\usepackage{epsfig}
\usepackage{epstopdf}
\usepackage{hyperref}
\hypersetup{colorlinks=true, pdfstartview=FitV, linkcolor=blue, citecolor=black, plainpages=false, pdfpagelabels=true, urlcolor=blue}
\usepackage[all]{hypcap}
\usepackage{amssymb}
\usepackage{amsmath}
\makegapedcells
\newcolumntype{R}[1]{>{\raggedleft\arraybackslash }b{#1}}
\newcolumntype{L}[1]{>{\raggedright\arraybackslash }b{#1}}
\newcolumntype{C}[1]{>{\centering\arraybackslash }b{#1}}
\usepackage{caption}

\usepackage{url}

\begin{document}
\mainmatter              % start of a contribution
\title{Multilayer Network Model of Movie Script}
%
%\titlerunning{Hamiltonian Mechanics}  % abbreviated title (for running head)
\author{Youssef Mourchid\inst{1} \and Benjamin Renoust\inst{2} \and
Hocine Cherifi\inst{3} \and Mohammed El Hassouni\inst{1, }\inst{4}}
\authorrunning{Youssef Mourchid et al.} % abbreviated author list (for running head)
\tocauthor{Youssef Mourchid, Benjamin Renoust, Hocine Cherifi and Mohammed El Hassouni}
\institute{LRIT-CNRST URAC 29, Rabat IT Center, Faculty of Sciences, Mohammed V University, Rabat, BP 1014, Morocco, {\small \email{youssefmour@gmail.com}}, \and Institute for Datability Science, Osaka University, Osaka, Japan, {\small \email{renoust@ids.osaka-u.ac.jp}},\and LE2I UMR 6306 CNRS, University of Burgundy, Dijon, France {\small \email{hocine.cherifi@u-bourgogne.fr}}, \and DESTEC, FLSH, Mohammed V University in Rabat, Morocco {\small \email{mohamed.elhassouni@gmail.com}}}

\maketitle              % typeset the title of the contribution

\begin{abstract}
Network models have been increasingly used in the past years to support summarization and analysis of narratives, such as famous TV series, books and news. Inspired by social network analysis, most of these models focus on the characters at play. The network model well captures all characters interactions, giving a broad picture of the narration's content. A few works went beyond by introducing additional semantic elements, always captured in a single layer network. In contrast, we introduce in this work a multilayer network model to capture more elements of the narration of a movie from its script: people, locations, and other semantic elements. This model enables new measures and insights on movies. We demonstrate this model on two very popular movies.
\end{abstract}

\section{Introduction}
\label{sec:introduction}

Whether narrated through books or movies, stories create their own universe, putting at play characters in a rich world. The intertwining of the different elements of a story forms the imaginary world that may catch the readers or viewers' full attention. Books leave to their readers the assembly of story elements glued with their own imagination for each to construct their own vision of the world depicted. 
Movies are very different in that, they serve viewers a fully constructed world for them only to enjoy with wonder. When some movie directors like to play with the viewers' progressive perception of their created universe, others really present rich imaginary worlds like in science-fiction movies. 
One could wish to understand if the interactions of the story elements can make a fingerprint of a story, characterizing a genre or a director.

The interactions of stories elements have often been captured through the use of network modelling \cite{park2012social,waumans2015topology,tan2014character,renoust2015social}. 
It has been used to support narration of a wide range of stories, from books \cite{waumans2015topology}, TV series \cite{tan2014character}, stories of news events \cite{renoust2015social}, and movies \cite{park2012social}, which make our focus. 
Networks are visuals objects that can not only be cleverly visualized to explicit the stories \cite{renoust2015social}, but also their structure can be interrogated in matter of \emph{topology} \cite{waumans2015topology,rital2005weighted}

These works are limited. Indeed, they mostly focus on a single facet of the stories, mainly the characters at play. 
Journalists often investigate a story by articulating the 5 W-questions: \emph{Who?}, \emph{Where?}, \emph{What?}, \emph{When?} and \emph{How/Why?} \cite{chen2009novel,kipling2010just}. 
The network analysis of stories attempts to answer \emph{How/Why?} by articulating the other 4 elements in a network. 
Since \emph{When?} is provided with data (tight with the narration), the previous works have mostly focused on \emph{Who?} in a single-layer network. 
We wish to introduce a more holistic approach on tackling also \emph{Where?} and \emph{What?} with a multi-layer network modeling of the stories.

Of course, the network may be manually created \cite{mish2016game}, but automated approaches may even scale to larger archives \cite{renoust2015social,waumans2015topology}. 
Since it is still hard to get all the necessary information from the video data itself \cite{demirkesen2008comparison,pastrana2006predicting}, we may rely on text analysis. Fortunately, in their very early stages of creation, movies are \emph{written} in form of scripts.
A script is usually extremely well structured, and contains all the necessary components to automatically analyze a movie \cite{jhala2008exploiting} (\emph{e.g.}  scenes, dialogues, characters, \emph{etc.}).

We propose in this paper to exploit this textual information to automatically extract the movie networks from the scripts. 
We contribute with a novel multilayer model of a movie script enabling to articulate characters, places, and themas. 
The multilayer network captures a richer structure of a movie. It completes the single character network analysis, and brings new topological analysis tools \cite{kivela2014multilayer}.

After discussing the related work in the next section, we introduce the proposed model in Section \ref{sec:model}. We describe the movie script processing in Section \ref{sec:processing}, before deploying the analysis in \ref{sec:usecases} on two popular science-fiction movies.
We finally conclude in Section \ref{sec:conclusion}

 \section{Introduction}
\label{sec:introduction}

Whether narrated through books or movies, stories create their own universe, putting at play characters in a rich world. The intertwining of the different elements of a story forms the imaginary world that may catch the readers or viewers' full attention. Books leave to their readers the assembly of story elements glued with their own imagination for each to construct their own vision of the world depicted. 
Movies are very different in that, they serve viewers a fully constructed world for them only to enjoy with wonder. When some movie directors like to play with the viewers' progressive perception of their created universe, others really present rich imaginary worlds like in science-fiction movies. 
One could wish to understand if the interactions of the story elements can make a fingerprint of a story, characterizing a genre or a director.

The interactions of stories elements have often been captured through the use of network modelling \cite{park2012social,waumans2015topology,tan2014character,renoust2015social}. 
It has been used to support narration of a wide range of stories, from books \cite{waumans2015topology}, TV series \cite{tan2014character}, stories of news events \cite{renoust2015social}, and movies \cite{park2012social}, which make our focus. 
Networks are visuals objects that can not only be cleverly visualized to explicit the stories \cite{renoust2015social}, but also their structure can be interrogated in matter of \emph{topology} \cite{waumans2015topology,rital2005weighted}

These works are limited. Indeed, they mostly focus on a single facet of the stories, mainly the characters at play. 
Journalists often investigate a story by articulating the 5 W-questions: \emph{Who?}, \emph{Where?}, \emph{What?}, \emph{When?} and \emph{How/Why?} \cite{chen2009novel,kipling2010just}. 
The network analysis of stories attempts to answer \emph{How/Why?} by articulating the other 4 elements in a network. 
Since \emph{When?} is provided with data (tight with the narration), the previous works have mostly focused on \emph{Who?} in a single-layer network. 
We wish to introduce a more holistic approach on tackling also \emph{Where?} and \emph{What?} with a multi-layer network modeling of the stories.

Of course, the network may be manually created \cite{mish2016game}, but automated approaches may even scale to larger archives \cite{renoust2015social,waumans2015topology}. 
Since it is still hard to get all the necessary information from the video data itself  
\cite{demirkesen2008comparison,pastrana2006predicting}, we may rely on text analysis. Fortunately, in their very early stages of creation, movies are \emph{written} in form of scripts.
A script is usually extremely well structured, and contains all the necessary components to automatically analyze a movie \cite{jhala2008exploiting} (\emph{e.g.}  scenes, dialogues, characters, \emph{etc.}).

We propose in this paper to exploit this textual information to automatically extract the movie networks from the scripts. 
We contribute with a novel multilayer model of a movie script enabling to articulate characters, places, and themas. 
The multilayer network captures a richer structure of a movie. It completes the single character network analysis, and brings new topological analysis tools \cite{kivela2014multilayer}.

After discussing the related work in the next section, we introduce the proposed model in Section \ref{sec:model}. We describe the movie script processing in Section \ref{sec:processing}, before deploying the analysis in \ref{sec:usecases} on two popular science-fiction movies.
We finally conclude in Section \ref{sec:conclusion}

\section{Related works}
\label{sec:relatedwork}

Networks have been explored to analyze videos data \cite{renoust2014entanglement} and shown efficient for topics and concept analysis \cite{kadushin2012understanding}. Particularly actor networks have been broadly analyzed from literature \cite{waumans2015topology}, from TV news videos \cite{renoust2015social}, and even a website is dedicated to the social analysis of Game of Thrones \cite{mish2016game}. 

Many techniques have been proposed to analyze movies using a graph structure. Based on story structure analysis, they support the inference of the meaning of interactions between movie objects or scenes through structural analysis. These approaches consist of two major steps: The first one is the construction of the story model, and the second one is the extraction of the information from the model. Once the network that connects entities such as characters, scenes or shots is constructed, semantic information is extracted from this network.

Yeung \emph{et al.} \cite{yeung1996extracting} proposed a scene transition graph and an analysis method for movie browsing and navigation. Each node in the scene transition graph denotes a cluster of shots. There is a link between two nodes $i$ and $j$ if a shot represented by node $i$ immediately precedes any shots represented by node $j$. This approach builds a network with interactions between shots and then analyzes it in order to extract the story units of scenes. This model uses only a hierarchical clustering of shots with visual primitives for browsing, but it cannot retrieve movie story elements such as actors, scenes, and dialogues.

Jung \emph{et al.} \cite{jung2004narrative} proposed a narrative structure graph to summarize a movie. The graph is composed of scene nodes which are narrative elements with character interactions, and connections between scenes decided by editorial relations. Using only scenes to construct a narrative structure graph for movie summarization is not sufficient. Indeed, story elements such as major characters and their interactions cannot be retrieved from this graph. 

Tan \emph{et al.} \cite{tan2014character} proposed an analysis of the character networks in two science fiction television series. These networks are constructed based on the scene co-occurrence between characters to indicate the presence of a connection. Global network topological measures such as the average path length, graph density, network diameter, average degree, are computed and found to be similar between the two series. Furthermore, various node centrality scores are computed and used to reflect on the interplay between the central characters and the overall narrative.

Some studies have been conducted to apply social network analysis (SNA) for movie story analysis. RoleNet \cite{weng2009rolenet} is an SNA-based approach that was proposed to analyze movie stories. It can identify automatically leading roles and corresponding communities by investigating the social interactions between characters using a weighted graph where nodes represent characters, and edges represent co-appearance relationships, \emph{i.e.} two characters appear in the scene. Edge weight represents the number of co-appearances of two characters in the same scene. However, using only scenes to model the movie story is not enough: some scenes may be very long, and others are short. Using an additional source such as the dialog would be more adaptable than this assumption.

Character-net \cite{park2012social} is another interesting work that proposes a story-based movie analysis method via social network analysis. While RoleNet uses co-appearance as relationship between characters, Character-net utilizes dialog. Edges are weighted by the quantity of dialogue exchanged by the characters. Once the weighted network is built, characters are classified according to their degree centrality value as major, minor or extra role. Finally, the classification result is used to detect the movie sequences through clique clustering, and major and minor role clustering.  

All of these works focus mostly  on a single facet of the stories, either the characters or the scenes. We propose to extract more elements of the narration of a movie such as characters, locations and themas from its scripts. Through a multilayer network, all the extracted elements with their interactions are grouped together in order to form the story of the movie. By articulating all these elements together, the network enables us to respond to the main questions about the story: Who (W1), Where (W2) and What (W3). It allows also to investigate the relative importance of the various types of nodes.

\section{Modeling Stories with a Multilayer Network}
\label{sec:model}

In investigative disciplines, the \emph{Four Ws} ( \emph{Who?}, \emph{Where?}, \emph{What?}, \emph{When?}) \cite{kipling2010just} are the fundamental questions that constitute the formula describing a complete story \cite{flint1917newspaper}. 
Inferring \emph{How/Why?} can be done while articulating the other \emph{Ws} making them essential bricks of analysis:

Given our context of movie understanding, we reformulate the four \emph{Ws} as follows: 
\begin{itemize}
    \item \emph{Who?} refers to the \textbf{characters} of a movie; 
    \item \emph{Where?} refers to the \textbf{locations} the action of a movie takes place; 
    \item \emph{What?} refers to the \textbf{subjects} the movie talks about. 
    \item \emph{When?} refers to the \textbf{time} guiding the succession of actions taught by the movie. 
\end{itemize}
Except for \emph{time}, each of answers to the \emph{Ws} -- \emph{characters}, \emph{locations}, and \emph{subjects} -- form the entities grounding our study.

Our goal is to help formulate movie understanding by articulating these four \emph{Ws}. 
A  multilayer graph model is used to put these elements together in relationship to form a story. 
This graph  is made of three node layers in order to represent each type of entity (characters, locations, and subjects -- represented by \emph{keywords}) with multiple relationships between them. 

\begin{figure}[htbp]
\centering
\includegraphics[width=10cm,height=5cm]{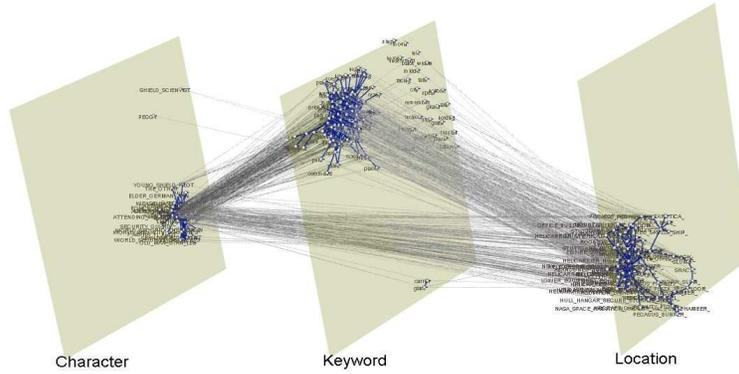}
\caption{Multilayer network extracted from Avengers script with the three layers of \emph{Character}, \emph{Location} \emph{Keyword} and their interactions (realized with MuxViz \cite{domenico2014multilayer})}
\label{fig:model}
\end{figure}

We model two main classes of relationships: intra-layer relationships  between nodes of a same category (\emph{e.g.} two people conversing); 
and inter-layer relationships between nodes of different categories (\emph{e.g.} a person being at a specific place); 
Put together, the multiple families of nodes and edges form a multilayer graph as illustrated in Fig.\ref{fig:model}.

We now define our multilayer graph $\mathbb G = (\mathbb V, \mathbb E)$ such that:

\begin{itemize}
\item $V_C \subseteq \mathbb V$ represents the set of characters $c \in V_C$,
\item $V_L \subseteq \mathbb V$ represents the set of locations $l \in V_L$,
\item $V_K \subseteq \mathbb V$ represents the set of keywords $k \in V_K$.
\end{itemize}

The different families of relationships can then be defined as:
\begin{itemize}
\item $e \in E_{CC} \subseteq \mathbb E$ between two characters such that $e=(c_i, c_j) \in V_C^2$, when a character $c_i \in V_C$ is conversing with another character $c_j \in V_C$.

\item $e \in E_{LL} \subseteq \mathbb E$ between two locations such that $e=(l_i, l_j) \in V_L^2$, when there is a temporal transition from one location $l_i \in V_L$ to the other $l_j \in V_L$.

\item $e \in E_{KK} \subseteq \mathbb E$ between two keywords such that $e=(k_i, k_j) \in V_K^2$, when $k_i \in V_K$ and $k_j \in V_K$ belong to the same subject.

\item $e \in E_{CK} \subseteq \mathbb E$ between a character and keyword such that $e=(c_i, k_j) \in V_C \times V_K$, when the keyword $k_j \in V_K$ is pronounced by the character $c_i \in V_C$. 

\item $e \in E_{CL} \subseteq \mathbb E$ between a character and a location such that $e=(c_i, l_j)  \in V_C \times V_L$, when a character $c_i \in V_C$ is present in location $l_j \in V_L$.

\item $e \in E_{KL} \subseteq \mathbb E$ between a keyword and a location such that $e=(k_i, l_j)  \in V_K \times V_K$, when a keyword $k_i \in V_K$ is mentioned in a conversation taking place in the location $l_j \in V_L$.
\end{itemize}

Note that for the sake of simplicity, we do not consider edge direction and weight. Furthermore, as we do not intend to study the network dynamics, time is not taken into account. However, time supports everything: the existence of a node or an edge is defined upon time, unrolled by the order of movie scenes.

As a shortcut, we can now refer to subgraphs by only considering one layer of links and its induced subgraph: 
\begin{itemize}
\item $G_{CC} = (V_C, E_{CC}) \subseteq \mathbb G$ refers to the subgraph of character interaction;
\item $G_{LL} = (V_L, E_{LL}) \subseteq \mathbb G$ refers to the subgraph of location transitions;
\item $G_{KK} = (V_K, E_{KK}) \subseteq \mathbb G$ refers to the subgraph of keyword co-occurrence;
\item $G_{CK} = (V_C \cup V_K, E_{CK}) \subseteq \mathbb G$ refers to the subgraph of characters speaking keywords;
\item $G_{CL} = (V_C \cup V_L, E_{CL}) \subseteq \mathbb G$ refers to the subgraph of characters standing at locations;
\item $G_{KL} = (V_K \cup V_L, E_{KK}) \subseteq \mathbb G$ refers to the subgraph of keywords mentioned at locations.
\end{itemize}

Now that we have set the model, we need to extract its elements from the script. This allows analyzing various topological properties of the network in order to gain a better understanding of the story.

\section{Extracting the Multilayer Network from the script}
\label{sec:processing}

We now describe the methodology used to build the network of the movie. Fortunately, most feature-length movies are produced using scripts that happen to be very well-structured textual documents \cite{jhala2008exploiting}. They organize the film in scenes, each placing characters in locations, describing their context, the actions, and most importantly the characters dialogues. As the script brings at play numerous but ambiguous notions, we first give essential information about its structure and a few important definitions. Then, we present the methodology used  to extract the various entities and interactions. Finally, we explain how to build the network based on this information. 

\subsection{Script Structure and Definitions}

\begin{figure}[htbp]
\centering
\includegraphics[width=0.7\linewidth]{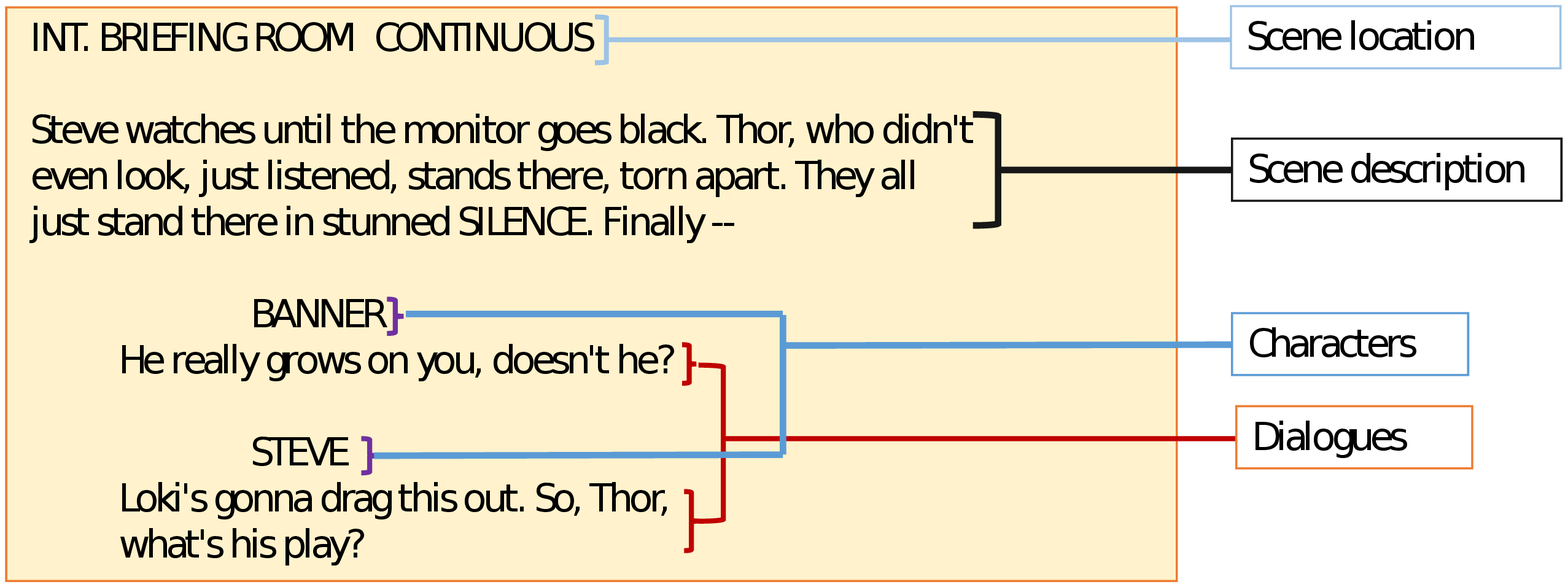}
\caption{Snippets of a resource script describing the movie \emph{The Avengers 2012}  \cite{avengers2012}, displaying different elements manipulated (characters, dialogues and locations).}
\label{fig:script}
\end{figure}

A script details all elements of a story: location and setting, characters with their situation and actions, and most importantly dialogues. The actual content of a script often follows a semi-regular format \cite{jhala2008exploiting} such as depicted in Fig. \ref{fig:script}.

A script usually starts with a heading describing the location and time of the scene. Specific keywords give important setting information (such as inside or outside scene) and character and key objects are often emphasized. The script then follows in a series of dialogues and setting descriptions. In order to remove any ambiguity, we first define the following dedicated glossary.

\begin{itemize}
\item Script: A text source of the movie which has descriptions about scenes, with setting and dialogues.
\item Scene: Chunk of a script, temporal unit of the movie. The collection of all scenes form the movie script.
\item Setting: The location a scene takes place, and its description.
\item Character: Denotes a person/animal/creature who is present in a scene, often impersonated by an actor.
\item Dialogues: A collection of utterances, what all characters say during a scene.
\item Utterance: An uninterrupted block of a dialogue pronounced by one character.
\item Conversation: A continuous series of utterances between two characters.
\item Speaker: A character who pronounced an utterance.
\item Description: A script block which describes the setting.
\item Location: Where does a scene take place, or mentioned by a character.
\item Keyword: Most relevant information from an utterance, often representative of its topic.

\end{itemize}

\subsection{Script Preprocessing}

\begin{figure}[htbp]
\centering
\includegraphics[width=0.9\linewidth, height=5.5cm]{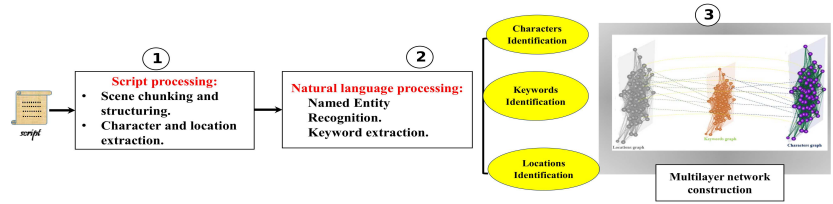}
\caption{Schematic view of the model construction process.}
\label{fig:methodology}
\end{figure}

Figure \ref{fig:methodology} illustrates the script processing pipeline. Although this process is language-dependant, we restrict our attention to scripts written in English. Whatever, generally, script content follows a semi-regular format as shown in Fig.\ref{fig:script}.

\textbf{Scene chunking: }

The first task is to chunk the script into scenes. Indeed, they are the main subdivisions of a movie, and consequently the main unit of analysis. To each scene is attached a set (actions, location and characters). Each scene starts with a technical description line written in capital letters, that establishes the physical context of the action that follows. It starts by indicating whether a scene takes place inside or outside (\emph{INT} or \emph{EXT}), the name of the location, and can potentially specify the time of day (e.g. \emph{DAY} or \emph{NIGHT}). These are markers we detect to chunk the script into scenes.  

\textbf{Scene structure: }
Sets are attached to locations which are always included in the scene header, that we can easily parse. Important people and key objects are usually highlighted in capital letters that we harvest while analyzing the text. The rest of a scene is made of dialogue and description. Character names and their actions are always depicted before the actual dialogue lines. A line indent also helps to identify characters and dialogue parts in contrast to scene description.

By structuring each scene into its list of description and dialogues, we can harvest set locations and utterance speakers. We can then identify scene conversations and characters present at a set. Specific descriptions can then be associated to locations, and dialogues to characters.

\subsection{Text processing}

The next step is to process the actual text content that is already associated either to a set location, or a to speaker. To do so, we use natural language processing tools to extract named entities and keywords. 

\textbf{Named Entity Recognition: }
Named Entity Recognition (NER) tag important words identified in a text content (such as people, organizations, cities, \emph{etc.}). We use the spaCy library \cite{al2017choosing} for its efficiency. We apply NER to scene description blocks and discard irrelevant categories such as quantities, ordinals, money \emph{etc.}. Because many words may end up mislabelled (especially due to the ambiguous context of a sci-fi movie), we manually curate the resulting list of words. We combine the categories to three larger categories: characters, location, and keywords, and we assign a unique class for ambiguous names referring to the same concept (\emph{i.e.} $\{TONY, STARK, IRON MAN\}\xrightarrow{}TONY STARK$). NER mainly helps us identifying characters present at a scene who are pronouncing utterances in conversations.

\textbf{Keyword extraction: }
Keywords are identified from dialogues. We measure the relevance of keywords using three methods: TF-IDF \cite{li2007keyword}, LDA \cite{blei2003latent} and Word2Vec \cite{yuepeng2015keyword}. Script texts are made of short sentences (even shorter after stop-words removal), so Word2Vec and TF-IDF render either too few words or too much of words without semantic content. Therefore, we mainly rely on LDA, which bring the best trade-off, and manually curate the resulting words by removing the semantic-less keywords (such as \emph{can}, \emph{have}, and so on).

\subsection{Network construction}

As a result of the previous two steps, for each scene, location and their description, characters, utterances and their speakers , and keywords extracted from description and utterances are identified. We now use this structure to construct the multilayer graph.

Let us revisit our investigative questions in the context of a scene: \emph{Who is involved in a scene?} may be tackled by characters, \emph{Where does a scene take place?} is identified through locations, and \emph{What is a scene about?} is identified through keywords. These three families of entities naturally form the three categories of nodes, respectively $V_C$, $V_L$ and $V_K$.

We now wish to infer the relationships we described in Section \ref{sec:model}. 
Two characters $c_i$, $c_j$ can be connected when they participate in a same conversation, hence forming an edge $e_{c_i, c_j} \in E_{CC}$. We connect two locations $e_{l_i, l_j} \in E_{LL}$ when there is a temporal transition between the locations $l_i$ and $l_j$ (analogous to geographical proximity), \emph{i.e.} following the succession of two scenes. Keywords $k_i$, $k_j$ co-occurring in a same conversation create an edge $e_{k_i, k_j} \in E_{KK}$.

Using the structure, extracted from the script, we can add additional links between categories. An edge $e_{c_i,l_j} \in E_{CL}$ associates a character $c_i$ with a location $l_j$ when the character $c_i$ appears in a scene taking place at location $l_j$. When a character $c_i$ speaks an utterance in a conversion, for each keyword $k_j$ that is detected in this utterance, we will create an edge $e_{c_i,k_j} \in E_{CK}$. Finally, we can associate the keywords $k_i$ extracted in conversation placed in a location $l_j$ to form the edge $e_{k_i,l_j} \in E_{KL}$.
A resulting graph combining all layers is visualized in Fig.\ref{fig:model}.

\section{Analyzing the Multilayer Network}
\label{sec:usecases}

\subsection{Data and Method}

We now wish to apply our methodology onto two popular science-fiction movie scripts, \emph{Star Wars: Episode IV-A New Hope} \cite{starwars1977episode} (hereafter \emph{SW}) and \emph{Marvel\'s the Avengers} \cite{avengers2012} (hereafter \emph{AV}). \emph{Star Wars}, 1977, is iconic of the science-fiction genre. It tells the story of a young man awaking to the Force (a religion as much as a source of power) in an epic battle between the Light (the Jedi) and Dark (the Empire) side of the Force dominating a galaxy. \emph{The Avengers}, 2012, puts at play a team of superheroes (the Avengers) from the Marvel universe against an extraterrestrial villain trying to invade Earth in search for the Tesseract, a powerful artifact.

the scripts are collected  from ({\small{\url{www.imsdb.com}}}), an online repository database made available by filmmakers. Once the graphs are extracted for each movie, we first investigate their global topological properties. Then, we investigate the node  \emph{influence} using classical centrality measures (Degree, Betweenness, Eigenvector) and a measure recently introduced that combine both local and global features (Influence Score) \cite{bioglio2017movie}.

\subsection{Basic topological properties}
Basic topological properties of the networks are reported in Table \ref{tab:statistics}. Note that the character and the location layers are made of a single connected component. The keyword layers have a few isolated nodes. We report the statistics of the "giant" component in this case. The multilayer networks of both movies are quite small. The Character layers are very dense as compared to location and keyword. Indeed, almost all characters have been talking together. In addition, all layers have a high diameter, especially for the location layer. This is due to a few temporal transitions between not very frequent locations that introduce long paths. In both movies, the clustering coefficient is very high for the character layer. The keywords layers are also well-clustered. There is no triangle in the inter-layer interactions (as in $G_{CL}$, $G_{CK}$ and $G_{KL}$) because those are bipartite graphs linking two sets of objects. We can observe that the multilayer networks are assortative. Indeed, major objects in the movie, especially characters and keywords tend to connect together. For example, “tesseract” and “Hellicarrier bridge” which are major keywords tend to be connected with “Nick Fury” a main character in the movie. Both multilayer networks exhibit a small average shortest path.

\begin{table}[htbp]

\begin{center}
    
\resizebox{0.8\textwidth}{!}{%
\begin{tabular}{c|c|c|c|c|c|c|c|c|c|c|c|c|c|c|}
\cline{2-15}
 & \multicolumn{7}{c|}{\textbf{Avengers}} & \multicolumn{7}{c|}{\textbf{Star}} \\ \cline{2-15} 
 & \textbf{V} & \textbf{E} & \textbf{$\rho$} & \textbf{$d$} & \textbf{$C$} & \textbf{$\tau$} & \textbf{$l_G$} & \textbf{V} & \textbf{E} & \textbf{$\rho$} & \textbf{$d$} & \textbf{$C$} & \textbf{$\tau$} & \textbf{$l_G$} \\ \hline
\multicolumn{1}{|c|}{\textbf{$G$}} & 187 & 1391 & 0,06 & 5 & 0,53 & 0,5 & 2,42 & 269 & 2586 & 0,06 & 6 & 0,56 & 0,52 & 2,61 \\ \hline
\multicolumn{1}{|c|}{\textbf{$G_{CC}$}} & 38 & 276 & 0,46 & 4 & 0,78 & 0,18 & 2.08 & 62 & 650 & 0,32 & 4 & 0,84 & 0,07 & 2.02 \\ \hline
\multicolumn{1}{|c|}{\textbf{$G_{KK}$}} & 81 & 221 & 0,07 & 7 & 0,49 & 0,05 & 2.83 & 74 & 701 & 0,08 & 7 & 0,57 & $2*10^{-2}$ & 3.46 \\ \hline
\multicolumn{1}{|c|}{\textbf{$G_{LL}$}} & 68 & 240 & 0,07 & 9 & 0,29 & 0,03 & 3.90 & 133 & 566 & 0,05 & 10 & 0,24 & 0,17 & 3.81 \\ \hline
\multicolumn{1}{|c|}{\textbf{$G_{CL}$}} & 81 & 159 & 0,04 & 6 & 0 & -0,01 & 3.36 & 143 & 316 & 0,04 & 7 & 0 & 0,07 & 3.24 \\ \hline
\multicolumn{1}{|c|}{\textbf{$G_{CK}$}} & 96 & 228 & 0,07 & 6 & 0 & -0,02 & 2.78 & 85 & 148 & 0,06 & 5 & 0 & -0,1 & 2.91 \\ \hline
\multicolumn{1}{|c|}{\textbf{$G_{KL}$}} & 115 & 267 & 0,04 & 6 & 0 & -0,02 & 3.14 & 96 & 205 & 0,06 & 10 & 0 & -0,02 & 3.27 \\ \hline
\end{tabular}%
}

\captionof{table}{Global topological properties of the various networks: number of nodes (V), number of edges (E), Density ($\rho$), Diameter ($d$), Average Clustering Coefficient (C), Assortativity Coefficient ($\tau$), and Average shortest path ($l_G$).}

\label{tab:statistics}

\end{center}
\end{table}

\subsection{Node influence}

We report in Table \ref{tab:staravenger} the top 5 nodes sorted by their centrality score computed in the three layers considered independantly for \emph{SW} and \emph{AV}. 

\textbf{Ranking characters:} All top characters are the main ones in \emph{SW} with \emph{Luke Skywalker} dominating all centralities. \emph{Vader}, \emph{Leia} and \emph{Han Solo} are accompanying the main character during all the movie. \emph{Leia} is the princess that is saved leading the rebels. \emph{Ben} the captain of the Millennium Falcon, work together with \emph{C-3PO} to rescue the beautiful princess. \emph{Darth Vader} is the main villain. In \emph{AV}, the top characters play an important role in the movie and the Marvel universe. Note that except for three characters they are superheroes (, also having their own movie (including the top-billed and occurring one, \emph{Tony Stark}). \emph{Nick Fury}, the top influential character is basically the liaison between all superheroes, which team-up with \emph{Tony Stark}, \emph{Captain America} and \emph{Natasha} against \emph{Loki} the last of our rank, which is the main villain.

\textbf{Ranking keywords:} In \emph{SW}, there is an extremely uneven distribution in the keywords centralities. In the top rankings we still find key concepts such as the \emph{Empire}, \emph{Jedi}, \emph{side}. The name of \emph{Luke} the main character is notably the most occurring by far. In \emph{AV}, the top keyword overall is the \emph{Tesseract}, which is the object motivating the villains to attack. The following keywords explain why so: \emph{world} and \emph{control} supports that this object is a powerful source which can control the world. \emph{Phil Coulson} is the one send by \emph{Nick fury} to find the heroes he needs to oppose Loki with the power of \emph{Tesseract}.

\textbf{Ranking locations:} In \emph{SW}, the top location is the \emph{Space}, where almost all the scenes take place.  \emph{Surface of the Death Star} -- in which another major part of the action takes place -- destroys it. The \emph{ Luke’s X-wing Fighter Cockpit} is Luke's ship which helps him in the war against \emph {Darth Vader’s Cockpit}. In \emph{AV}, the top three locations are the places where all main characters regroup or fight, and in which most action occurs (note that the \emph{Helicarrier Bridge} and \emph{Helicarrier Detention Section} are two places of the \emph{Quinjet}, which is the vehicle transporting the main characters everywhere). \emph{Stark Tower} is \emph{Tony Stark's} home which is another key place in the movie .\emph{Sky} is a transitory scene. \emph{Manhattan} is the city where the fight begins between superheroes and Loki.
We can observe that for both movie, the Influence score ranking is globally well correlated with the fraction of appearance of the objects per scene (Characters, Keywords, Locations).

\begin{table}[htbp]

\resizebox{1.\textwidth}{!}{%
\begin{tabular}{c|c|c|c|c|c|c|c|c|c|c|c|c|c|c|c|}
\cline{2-16}
 & \multicolumn{5}{c|}{CHARACTERS} & \multicolumn{5}{c|}{KEYWORDS} & \multicolumn{5}{c|}{LOCATIONS} \\ \cline{2-16} 
 & D & B & Ei & I.S & O & D & B & Ei & I.S & O & D & B & Ei & I.S & O \\ \hline
\multicolumn{1}{|c|}{\multirow{5}{*}{SW}} & Luke & Luke & Luke & Luke & 89 & Empire & Side & Luke & Empire & 7 & S & S & Sds & S & 31 \\ \cline{2-16} 
\multicolumn{1}{|c|}{} & C-3PO & C-3PO & Han & C-3PO & 41 & Luke & Empire & Side & Luke & 180 & Sds & Mfc & Lxfc & Sds & 29 \\ \cline{2-16} 
\multicolumn{1}{|c|}{} & Han & Vader & C-3PO & Han & 42 & Jedi & Sandpeople & Jedi & Side & 61 & Lxfc & A & Dvc & Lxfc & 36 \\ \cline{2-16} 
\multicolumn{1}{|c|}{} & Vader & Han & Leia & Vader & 33 & Side & Luke & System & Long & 38 & Sads & Dsh & Sads & Sads & 26 \\ \cline{2-16} 
\multicolumn{1}{|c|}{} & Leia & Red leader & Ben & Leia & 24 & System & Stay & Back & Jedi & 11 & Mfc & Td & Rlc & Dvc & 21 \\ \hline
\multicolumn{1}{|c|}{\multirow{5}{*}{AV}} & N.Fury & N.Fury & C.America & N.Fury & 36 & Tesseract & Tesseract & Tesseract & Tesseract & 31 & Hb & Hb & Hb & Hb & 19 \\ \cline{2-16} 
\multicolumn{1}{|c|}{} & Tony & Banner & Tony & Tony & 48 & P.Coulson & P.Coulson & P.Coulson & P.Coulson & 17 & M & Hds & M & Hds & 10 \\ \cline{2-16} 
\multicolumn{1}{|c|}{} & C.America & C.America & N.Fury & C.America & 69 & World & Tony & World & World & 17 & St & St & St & St & 11 \\ \cline{2-16} 
\multicolumn{1}{|c|}{} & Natasha & Loki & Natasha & Natasha & 36 & Tony & World & Tony & Tony & 2 & Sk & Sk & Sk & Sk & 9 \\ \cline{2-16} 
\multicolumn{1}{|c|}{} & Banner & Natasha & Banner & Loki & 35 & Control & Control & Control & Control & 17 & Hds & M & Hds & M & 13 \\ \hline
\end{tabular}%
}
\captionof{table}{First 5 nodes sorted by relevance according to their centrality score: Degree(D), Betweenness(B), Eigencentrality(Ei), Influence score(I.S). (0) is the fraction of scenes where the objects appears. In the location layer abbreviations mean S:Space;A:Alderaan;Dsh: Death Star Hallway; Td: Tatooine Desert; Dvc: Darth Vader's Cockpit; Sds: Surface of The Death Star; Lxfc: Luke's X-wing Fighter Cockpit; Sds:Space Around The Death Star; Mfc:Millennium Falcon Cockpit; Hb:Helicarrier Bridge; M:Manhattan; St:Sark Tower; Sk:Sky; Hds:Helicarrier Detention Section.}
\label{tab:staravenger}

\end{table}

\textbf{Multilayer ranking:} 
Table \ref{tab:multi} reports the centrality scores computed in the multilayer networks. The top nodes are mainly the main characters of the movie. However, major Locations and keywords are also in the top ranking list.  We can observe that while degree centrality tends to bring forward characters, betweenness and eigencentrality measures tend to bring forward locations.

To summarize, these results show that even a very succinct topological analysis of the multilayer networks can give a pretty good idea about the movie content.

\begin{table}[htbp]

\resizebox{\textwidth}{!}{%
\begin{tabular}{|c|c|c|c|c|c|c|c|c|c|c|c|c|c|}
\hline
\multirow{4}{*}{AV} & D & N.Fury & Natasha & Tony & Banner & C.America & Loki & Thor & P.Coulson & Barton & Tesseract & Hb & Barton \\ \cline{2-14} 
 & B & N.Fury & Banner & Natasha & Hb & Loki & Tony & Thor & C.America & Bg & Bl & Tesseract & Bs \\ \cline{2-14} 
 & Ei & C.America & Tony & N.Fury & Natasha & Thor & Banner & Loki & P.Coulson & Bl & Bs & Loki & Hb \\ \cline{2-14} 
 & I.S & N.Fury & Natasha & Tony & Banner & C.America & Loki & Thor & P.Coulson & Barton & Tesseract & Hb & Barton \\ \hline
\multirow{4}{*}{SW} & D & Luke & C-3PO & Han & Ben & Leia & Vader & Biggs & Tarkin & R.Leader & S & Sds & Lxfc \\ \cline{2-14} 
 & B & Luke & C-3PO & S & Han & A & Leia & Vader & Ben & Sds & Biggs & Mfc & Td \\ \cline{2-14} 
 & I.S & Luke & C-3PO & Han & Ben & Leia & Vader & Biggs & Tarkin & R.Leader & S & Sds & Lxfc \\ \hline
\end{tabular}%
}

\captionof{table}{First 12 nodes sorted by centrality scores for the multilayer networks. In the location layer abbreviation mean, Bg:Brooklyn Gym; Bl:Banner's Lab; Bs:Bridge Street; Mfc:Millenium Falcon Cockpit; Td:Tatooine Desert; Tmes:Tatooine Mos Eisley Street.
}
\label{tab:multi}

\end{table}

\section{Conclusion}
\label{sec:conclusion}
In this paper, we introduce a multilayer model that relates  \emph{characters}, \emph{locations} and \emph{keywords} in movies. This model is much more informative than single layer networks which usually concentrate on \emph{characters} or \emph{scenes}. We also propose a methodology to extract the networks elements from the script. We have deployed the model on two popular movies. At the moment, the processing requires to clean the results. In order to get a fully automatic network extractor, we need to remove any ambiguity by including multimedia elements, such as scene segmentation, face tracking, and subtitle information. Due to space constraint, we reported the results of a brief analysis of the extracted networks that confirmed the effectiveness of the model.  Note that much more information can be gained by a deeper topological analysis.To be the less ambiguous possible, we have considered only primitive interactions that are easy to interpret, but we could easily project different views. The model can be extended to derive a co-occurrence network of characters in the same location, a directed network of of conversations, or mention of characters, \emph{etc.}.Future work implies deploying our tool on larger collections, such as the whole Star Wars series, or even a larger collection of movies. We plan to use the multilayer network model to characterize movie period, genres, or directors, and even correlate with acting careers from public databases such as IMDB.

\bibliographystyle{spmpsci}

\end{document}